\documentstyle[prl,aps,epsf]{revtex}

\begin{document}
\draft

\title{On the ground state of a completely filled lowest Landau level in two dimensions}
\author{S. A. Mikhailov}
\address{
Institute for Theoretical Physics, University of Regensburg, 93040 Regensburg, Germany
}
\date{\today}

\maketitle

\begin{abstract}
There exists a widely believed opinion, that the many-body ground state of a two-dimensional electron system at a completely filled lowest Landau level (the filling factor $\nu=1$) is described by the so-called Hartree-Fock wave function, and that this solution is the unique, exact eigenstate of the system at $\nu=1$. I show that this opinion is erroneous, construct an infinite number of other variational many-body wave functions, and discuss the properties of a few states which have the energy substantially lower than the energy of the Hartree-Fock state.
\end{abstract}

The nature of the ground state of a system of two-dimensional (2D) interacting electrons in strong magnetic fields ${\bf B}$ was a subject of intensive investigations in the past years~\cite{Fukuyama79,Yoshioka79,Yoshioka83b,Maki83,Yoshioka83a,Laughlin83,Haldane83,Yoshioka84b,Lam84,Levesque84,Haldane85,MacDonald85,Kivelson86,Morf86,Jain89a,Halperin93,Aleiner95,Koulakov96,Fogler96,Moessner96,Fogler97,Jungwirth00,MacDonald00,Bychkov81}. A great deal of attention was given to a partially filled lowest ($\nu<1$)~\cite{Fukuyama79,Yoshioka79,Yoshioka83b,Maki83,Yoshioka83a,Laughlin83,Haldane83,Yoshioka84b,Lam84,Levesque84,Haldane85,MacDonald85,Kivelson86,Morf86,Jain89a,Halperin93} or higher Landau levels ($\nu\gg 1$)~\cite{Aleiner95,Koulakov96,Fogler96,Moessner96,Fogler97,Jungwirth00,MacDonald00}. The case of a completely filled lowest Landau level $\nu=1$ was not adequately investigated in the literature. The only trial many-body wave function for this $\nu$, the so-called Hartree-Fock solution, 
\begin{equation}
\Psi_{HF}^{[N]}= \frac 1{\sqrt{N!}}\det|\psi_{L_j}({\bf r}_i)|,{\ \ }L_j=0,1,\dots,N-1,
\label{HF}
\end{equation}
was proposed in Ref.~\cite{Bychkov81}. This is a Slater determinant, constructed from the lowest-Landau-level single-particle states
\begin{equation}
\psi_L({\bf r})= \frac{(z^*)^L}{\lambda\sqrt{\pi L!}}\exp(-zz^*/2),
\label{spst}
\end{equation}
and it is assumed that in the $N$-electron system the states $L=0$ to $L=N-1$ are occupied by electrons. Here $z=(x+iy)/\lambda$ is a complex coordinate of an electron, $\lambda^2=2l^2=2\hbar c/eB$, $l$ is the magnetic length, $L=0,1,2,\dots$ is the angular momentum quantum number, and $\hbar$, $c$, and $e$ are the Planck constant, velocity of light and the electron charge, respectively. The wave function (\ref{HF}) is an eigenfunction of the kinetic energy operator, with the eigenenergy $E=N\hbar\omega_c/2$. It coincides with the Laughlin many-body wave function~\cite{Laughlin83} at $\nu=1$. The state (\ref{HF}) is characterized, in the thermodynamic limit, by a uniform 2D electron density $n_s({\bf r})=1/\pi\lambda^2$ and the energy per particle~\cite{Laughlin83}
\begin{equation}
\epsilon_{HF}=-\pi/2=-1.57080
\label{HFenergy}
\end{equation}
in the $B$-independent energy units $e^2\sqrt{n_s}$. 

There exists an opinion, that the Hartree-Fock many body wave function (\ref{HF}) is {\it the only one} possible solution of the problem at $\nu=1$. Although this opinion is incorrect, it seems to be widely believed. The aim of this Letter is to show that the Hartree-Fock many-body wave function is not the only one possible trial wave function for the ground state of a 2DES at $\nu=1$, and to demonstrate a number of other variational solutions of the many-body Schr\"odinger equation at $\nu=1$, which have the energy substantially lower than the energy of the Hartree-Fock state. 

I argue in three different manners.

The first argument is quite simple: in order to prove that the function (\ref{HF}) is not unique at $\nu=1$, it is sufficient to present another explicit example of a trial wave function. Consider for instance the Wigner crystal many-body wave function~\cite{Maki83},
\begin{equation}
\Psi_{WC}^{[N]}=\frac 1{\sqrt{N!}}\det|\chi_{L=0}({\bf r}_i,{\bf R}_j)|,
\label{WC}
\end{equation}
\begin{equation}
\chi_L({\bf r}_i,{\bf R}_j)=\psi_L({\bf r}_i-{\bf R}_j) e^{-i\pi{\bf r}_i\cdot({\bf B\times R}_j)/\phi_0}.
\label{chi-functions}
\end{equation}
Here $\phi_0$ is the flux quantum and ${\bf R}_j$ are points of a triangular lattice~\cite{Maki83}, distributed over the 2D plane with the average density $n_s$. This function depends on magnetic field and can be considered at $\nu=1$. One can easily see that the projection of the Wigner crystal wave function onto the Hartree-Fock one is not unity. Expand the function $\chi_0({\bf r},{\bf R})$ in a set of the lowest-Landau-level eigenstates $\psi_L({\bf r})$,
\begin{equation} 
\chi_0({\bf r},{\bf R})\equiv\frac 1{\sqrt{\pi}\lambda} e^{-zz^*/2-ZZ^*/2+z^*Z}= \frac {e^{-zz^*/2-ZZ^*/2}}{\sqrt{\pi}\lambda}\sum_{L=0}^\infty \frac{(z^*Z)^L}{L!}= \sum_{L=0}^\infty C_L(Z)\psi_L({\bf r}),
\end{equation}
where $C_L(Z)= Z^L\exp(-ZZ^*/2)/\sqrt{L!}$. Then, for the two-electron function (\ref{WC}) I get 
\begin{eqnarray}
\Psi_{WC}^{[N=2]}
&=& \frac 1{\sqrt{2!}}
\det
\left |\begin{array}{cc}
\chi_0({\bf r}_1,{\bf R}_1)&
\chi_0({\bf r}_1,{\bf R}_2)\\
\chi_0({\bf r}_2,{\bf R}_1)&
\chi_0({\bf r}_2,{\bf R}_2)
\end{array}
\right |\nonumber \\ 
&=&\sum_{L_1=0}^\infty \sum_{L_2=0}^\infty C_{L_1}(Z_1)C_{L_2}(Z_2)\frac 1{\sqrt{2!}}\det\left |\begin{array}{cc}
\psi_{L_1}({\bf r}_1)& 
\psi_{L_2}({\bf r}_1)\\
\psi_{L_1}({\bf r}_2)& 
\psi_{L_2}({\bf r}_2)
\end{array}
\right |\nonumber \\ 
&=&\sum_{L_1=0}^\infty \sum_{L_2>L_1}^\infty \det
\left |\begin{array}{cc}
C_{L_1}(Z_1)&
C_{L_2}(Z_1)\\
C_{L_1}(Z_2)&
C_{L_2}(Z_2)
\end{array}
\right | \frac 1{\sqrt{2!}}\det
\left |\begin{array}{cc}
\psi_{L_1}({\bf r}_1)& 
\psi_{L_2}({\bf r}_1)\\
\psi_{L_1}({\bf r}_2)& 
\psi_{L_2}({\bf r}_2)
\end{array}
\right | .
\end{eqnarray}
All the basis functions $\det |\psi_{L_j}({\bf r}_i)|/\sqrt{2!}$ here are orthonormal. One of them (with $L_1=0$ and $L_2=1$) is the Hartree-Fock function (\ref{HF}). The projection
\begin{equation}
P\equiv\frac{\left|\langle WC|HF\rangle\right|^2}
{\langle WC|WC\rangle
\langle HF|HF\rangle}=\frac{\left |\det|C_{L_j}(Z_i)|\right|^2_{L_1=0,L_2=1}}
{\sum_{L_1=0}^\infty \sum_{L_2>L1}^\infty \left |\det|C_{L_j}(Z_i)|\right|^2}
\label{projection}
\end{equation}
is evidently smaller than one. The same derivation can be easily performed at any $N$. In the thermodynamic limit $N\to\infty$ the projection (\ref{projection}) tends to zero, $P\to 0$. 

Arbitrarily varying the vectors ${\bf R}_j$ (square lattice, other types of the lattice, random distribution), as well as taking the functions 
\begin{equation}
\Psi_L^{[N]}=\frac 1{\sqrt{N!}}\det|\chi_{L}({\bf r}_i,{\bf R}_j)|,
\label{PsiL}
\end{equation}
with other angular momentum index $L$, one can easily get an infinite number of other explicit examples of many-body wave functions, different from (\ref{HF}). 

The second argument is based on a standard degenerate perturbation theory. Consider the problem of $N$ 2D electrons in a perpendicular magnetic field, in the presence of a neutralizing positive background, which has the form of a disk with the radius $R$ and the charge density $+en_s$. The electroneutrality condition requires that $\pi R^2n_s=N$. Assume that the total Coulomb energy of the system (electron-electron plus electron-background plus background-background interaction energy) can be considered as a perturbation. The ground state of the unperturbed problem is highly degenerate, therefore, one should use a degenerate perturbation theory. In the one-electron problem, one should search for a solution in the form
$\Psi^{[N=1]}({\bf r})= \sum_{L=0}^\infty C_L\psi_L({\bf r})$. 
This expansion 
contains all $L$-terms, from $L=0$ to $L=\infty$. In the $N$-electron problem, one should write, similarly, 
\begin{equation}
\Psi^{[N]}({\bf r}_1,\dots,{\bf r}_N)=\sum_{L_1,\dots,L_N}^\infty C_{L_1,\dots,L_N}
\frac 1{\sqrt{N!}}\det | \psi_{L_j}({\bf r}_i)|,
\label{PsiNexp}
\end{equation}
where, again, all $L_i$ vary from $L_i=0$ to $L_i=\infty$. The Hartree-Fock Slater determinant is the only one term in this expansion ($L_1=0$, $L_2=1$, $\dots$, $L_N=N-1$). Obviously, any variational solution of the many-body Schr\"odinger equation can be also searched for in the form of an arbitrary linear combination (\ref{PsiNexp}). Hence, the number of all possible trial wave functions (at any $\nu$) is not one but infinite. 

Finally (the third argument) I discuss the problem in the finite-cell geometry. In a number of papers (see e.g. Ref.~\cite{Yoshioka83a}), the problem of a single 2D electron in a perpendicular magnetic field was considered in a finite rectangular cell ($0\le x\le a$, $0\le y\le b$), with periodic boundary conditions at the boundaries of the cell. According to~\cite{Yoshioka83a}, the ``boundary condition requires that $ab/2\pi l^2$ be an integer $m$'', and ``there are $m$ different single-electron states in the cell''
\begin{equation}
\phi_j({\bf r})=\left(\frac 1{b\sqrt{\pi}l}\right)^{1/2}
\sum_{k=-\infty}^\infty\exp\left[i\frac{(X_j+ka)y}{l^2}-\frac{(X_j+ka-x)^2}{2l^2}\right],
\label{wf}
\end{equation}
where $1\le j\le m$ and $X_j=2\pi l^2j/b$. From this statement it follows that, at $\nu=1$ the number of electrons per cell is exactly equal to the number of single-particle states, and therefore there exists only one possible way to construct the many-body wave function. Consider this argument in some more detail.

Consider a rigorous mathematical formulation of the single-electron problem in a finite-cell geometry. The wave function sought should satisfy the differential equation 
\begin{equation}
\left[-\hbar^2 \partial^2_x+\left(-i\hbar \partial_y+eBx/c\right)^2\right]\phi(x,y)=2mE\phi(x,y),
\label{eq}
\end{equation}
in the area $0\le x\le a$, $0\le y\le b$, and the boundary conditions
\begin{equation}
\phi(x,y=0)=\phi(x,y=b),\ \partial_y\phi(x,y=0)=\partial_y\phi(x,y=b), {\ \ } {\rm at\ all\ \ } 0\le x\le a,
\label{bc1}
\end{equation}
\begin{equation}
\phi(x=0,y)=\phi(x=a,y),\ \partial_x\phi(x=0,y)=\partial_x\phi(x=a,y), {\ \ } {\rm at\ all\ \ } 0\le y\le b.
\label{bc3}
\end{equation}
Standard substitutions, $\phi(x,y)=\varphi(x)\exp(ik_yy)$, $\xi=x/l+k_yl$, $\epsilon=2E/\hbar\omega_c$, lead to the following equation for $\tilde\varphi(\xi)\equiv\varphi(x=l\xi-k_yl^2)$,
\begin{equation}
-\tilde\varphi^{\prime\prime}(\xi)+\xi^2\tilde\varphi(\xi)=\epsilon\tilde\varphi(\xi).
\label{de}
\end{equation}
The second-order differential equation (\ref{de}) has two independent solutions, e.g. 
\begin{equation}
\Phi_1(\xi,\epsilon)=e^{-\xi^2/2}\left[1+(1-\epsilon)\xi^2/2!+
(1-\epsilon)(5-\epsilon)\xi^4/4!+\dots\right]
\end{equation}
and
\begin{equation}
\Phi_2(\xi,\epsilon)=e^{-\xi^2/2}\left[\xi+(3-\epsilon)\xi^3/3!+
(3-\epsilon)(7-\epsilon)\xi^5/5!+\dots\right].
\end{equation}
The total solution of eq.~(\ref{eq}) is then written in the form
\begin{equation}
\phi(x,y)=e^{ik_yy}\left[C_1\Phi_1(x/l+k_yl,\epsilon)+C_2\Phi_2(x/l+k_yl,\epsilon)\right],
\end{equation}
with two arbitrary constants $C_1$ and $C_2$. The boundary condition (\ref{bc1}) requires that $k_y=2\pi m/b$ with integer $m$. If the second boundary condition was imposed at infinity [$\phi(x=\pm \infty,y) =0$] one would get a conventional solution of the problem, with $\epsilon_n=2n+1$, $n=0,1,\dots$, and Landau eigenfunctions. In the finite-cell geometry, the boundary conditions (\ref{bc3}) require that
\begin{equation}
\begin{array}{c}
\{\Phi_1(k_yl,\epsilon)-\Phi_1(a/l+k_yl,\epsilon)\}C_1+\{\Phi_2(k_yl,\epsilon)-\Phi_2(a/l+k_yl,\epsilon)\}C_2=0, \\
\{\Phi_1^\prime(k_yl,\epsilon)-\Phi_1^\prime(a/l+k_yl,\epsilon)\}C_1+\{\Phi_2^\prime(k_yl,\epsilon)-\Phi_2^\prime(a/l+k_yl,\epsilon)\}C_2=0.
\end{array}
\label{dispeq}
\end{equation}
Equations (\ref{dispeq}) determine the spectrum of eigenenergies $\epsilon=\epsilon_n(a/l,b/l)$, $n=0,1,\dots$, and eigenfunctions of the problem (they are different from the Landau ones). Obviously, the boundary conditions do not restrict the number of eigenstates, and the single-particle spectrum remains unlimited.

One can easily verify, by means of a direct substitution, that the wave functions (\ref{wf}), proposed in~\cite{Yoshioka83a}, {\it do not satisfy} the boundary condition (\ref{bc3}). Instead, they satisfy the boundary condition $\phi(x=a,y)=\exp(iay/l^2)\phi(x=0,y)$. Hence, the functions (\ref{wf}) are not the eigenfunctions of the boundary-value problem (\ref{eq})--(\ref{bc3}), and the conclusion of Ref.~\cite{Yoshioka83a} about a finite number of possible single-particle states in the considered problem is incorrect. 

Now I present results of calculations of the energy of a few trial wave functions, different from the Hartree-Fock solution (\ref{HF}). I consider the functions $\Psi_L$, eq.~(\ref{PsiL}), with $L=0$ to 5, and with a triangular lattice of points ${\bf R}_j$, uniformly distributed over the 2D plane with the average density $n_s$. The states $\Psi_L$ describe the properties of the system at all Landau-level filling factors $\nu\le 1$, continuously as a function of magnetic field. Like in~\cite{Mikhailov00a}, I calculate the Hartree energy exactly, in the thermodynamic limit, and the exchange-correlation energy approximately, at a finite number $N$ of lattice points. Thus calculated energy per particle $\epsilon_L(N)$ is the upper limit to the true energy $\epsilon_L$, $\epsilon_L<\epsilon_L(N)$~\cite{Mikhailov00a}. 

\begin{figure}
\begin{center}
\begin{math}
\epsfxsize=8.5cm
\epsffile{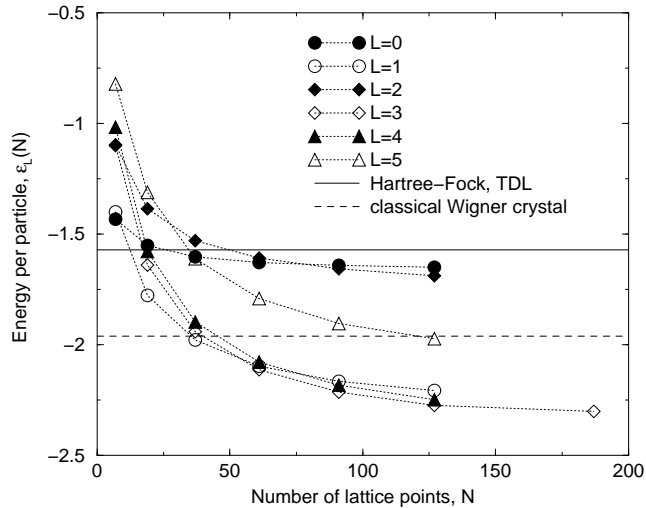}
\end{math}
\end{center}
\caption{Calculated energies $\epsilon_L(N)$ per particle of the states $\Psi_L$ (in units $e^2\sqrt{n_s}$), as a function of the number of (triangular) lattice points $N$, for the states $L=0$ to 5 at a completely filled lowest Landau level $\nu=1$. The Hartree contribution is calculated in the thermodynamic limit, the exchange-correlation contribution -- for a finite number $N$ of lattice points. For comparison, the energies of the Hartree-Fock state (the thermodynamic limit, TDL) and of the classical Wigner crystal \protect\cite{Bonsall77} are shown by solid and dashed lines respectively. }
\label{fig1}
\end{figure}

Figure \ref{fig1} exhibits the energy $\epsilon_L(N)$ of the states $\Psi_L$, at $\nu=1$, as a function of the number of lattice points $N$ involved in calculation of the exchange-correlation energy. {\it All} the states $\Psi_L$, $L=0,1,\dots,5$, have the energy lower than the Hartree-Fock one. Some of them ($L=1,3,4$ and 5) have the energy lower than the classical Wigner crystal~\cite{Bonsall77}. The energy of the state $\Psi_{L=3}$ estimated with $N=187$ lattice points, is more than 46\% 
lower than the energy of the Hartree-Fock state. All the considered states are characterized by a strong overlap of the neighbor single-particle wave functions and by a uniform 2D electron density $n_s({\bf r})=1/\pi\lambda^2$ at $\nu=1$ (see for example, Figure 3 from Ref.~\cite{Mikhailov00a}). Subsequent investigations of other trial wave functions (with other $L$ and/or different symmetries of the lattice ${\bf R}_j$) may lead to even lower estimates of the ground state energy of the 2DES at $\nu=1$. A huge variational freedom of the functions $\Psi_L$, which has been clearly demonstrated in this Letter, opens up wide possibilities to search for better approximations to the ground state, both at $\nu=1$ and at all other $\nu$.

This work was supported by the Graduiertenkolleg {\sl Komplexit\"at in Festk\"orpern}, University of Regensburg, Germany and the DFG-Sonderforschungsbereich 348 (Nanometer-Halbleiterbauelemente). I thank Ulrich R\"ossler, Nadejda Savostianova, Vladimir Volkov, Vladimir Sandomirsky, Oleg Pankratov, Robert Laughlin, Pawel Hawrylak and Allan MacDonald for discussions and useful comments.

\end{document}